# Comment to "Quantum Spin Hall Effect abd Topological Phase Transitio in HGTe Quantum Wells" by B. Andrei Bernevig, Taylor L. Hughes and Shou-Cheng Zhang Science 314, 1757 (2006).

Stanisław Krukowski[1], Paweł Kempisty[1], Paweł Strak[1] and Konrad Sakowski[1,2]

[1]Institute of High Pressure Physics, Polish Academy of Sciences, Sokołowska 29/37, 01-142 Warsaw, Poland

[2]University of Warsaw, Faculty of Mathematics, Informatics and Mechanics, Institute of Applied Mathematics and Mechanics, Banacha 2, 02-097 Warsaw, Poland

In the above paper (1), Bernevig, Hughes and Zhang(BHZ) present derivation of the topological surface state in the HgTe/CdTe mulitiquantum well system (MQWs). The model, is reduced version of the full eight state HgTe Hamiltonian published by Novik et al. (2). The 8 state space, modified by rejection of the spin-orbit decoupled states, was solved by BHZ as eigenvalue problem for HgTe/CdTe quantum wells (QWs) for several pairs of states, separately (1 - Suppl. online material). The equations they used, i.e. Eqs. 7 and 8 in (1) describe the wavefunctions dependence along z direction, perpendicular to the MQWs plane as:

$$Tf_1(z) + \sqrt{\frac{2}{3}}P(-i\partial_z)f_4(z) = Ef_1(z) \tag{1a}$$

$$\sqrt{\frac{2}{3}}P(-i\partial_z)f_1(z) + (U-V)f_4(z) = Ef_4(z) \tag{1b}$$

where $T = E_c - A\partial_z^2$, $U = E_v + \gamma_1\partial_z^2$, $U = -2\gamma_2\partial_z^2$ are the operators having different values in the well (HgTe) and in the barriers (CdTe). Mathematically these are the step functions for the z coordinate values corresponding to the well ($|z| < d/2$) and to the barriers ($|z| \geq d/2$), where $d$ is the well width. According to BHZ, the solutions $f_1(z)$ and $f_4(z)$ are odd and even functions with respect to reflection $z \leftrightarrow -z$, given as:

$$f_1(z) = \begin{cases} C_{CdTe}e^{\alpha z} & z < -d/2 \\ C_{HgTe}(e^{\delta z} + e^{-\delta z}) & |z| < d/2 \\ C_{CdTe}e^{-\alpha z} & z > d/2 \end{cases} \quad (2a)$$

$$f_4(z) = \begin{cases} V_{CdTe}e^{\alpha z} & z < -d/2 \\ V_{HgTe}(e^{\delta z} - e^{-\delta z}) & |z| < d/2 \\ -V_{CdTe}e^{-\alpha z} & z > d/2 \end{cases} \quad (2a)$$

The above ansatz was based on assumption of existence of interface states, in the HgTe well.

The Hamiltonian (Eq. 1) may be simplified and translated into the formulation used by Qi and Zhang (3), giving:

$$(M - B\partial_z^2)f_1(z) + A(-i\partial_z)f_4(z) = Ef_1(z) \quad (3a)$$

$$A(-i\partial_z)f_1(z) - (M - B\partial_z^2)f_4(z) = Ef_4(z) \quad (3b)$$

where $M = E_c = -E_v$, i.e. $E_c - E_v = 2M$, $A = \sqrt{\frac{2}{3}}P$ and $B = A = \gamma_1 - 2\gamma_2$. Therefore the only simplification is related to effective mass in conduction and valence band which are set both equal and isotropic. These equation has to be solved separately for the well and the barriers, giving the same eigenvalue $E$. The problem may be solved in the Fourier basis $f(z) = Ce^{ikz}$ which gives the solutions:

$$E_{\mp} = \mp\sqrt{(M + Bk^2)^2 + A^2k^2} \quad (4)$$

for electrons (+) and holes (-) respectively. The energy is divided into potential and kinetic contribution in the surrounding of the minimum, i.e.

$$E_e = E_{kin} + V_o = Fk^2 + V_o \quad (5)$$

where $V_o = M$ and $F = B + A^2/2M > 0$. The BHZ solution ansatz is obtained by the following substitution: $\alpha = ik = \sqrt{(E_e - V_o)/F}$ and $\delta = ik' = \sqrt{(E_e - V_o)/F}$. The kinetic energy of the quantum state consists of two contribution: barrier $E_{kin}(bar) = \int Fk^2 dz = -\int F\alpha^2 dz$ and the well $E_{kin}(well) = \int Fk^2 dz = -\int F\delta^2 dz$. Both contributions are negative: $E_{kin}(bar) < 0$ and $E_{kin}(well) < 0$. That is not allowed as the kinetic energy is positive definite. The only possible case is that state is the well is oscillatory, i.e. it is bulk state as the one given by Eq. 13 (1 – Suppl.). In this case the negative contribution to kinetic energy of exponential decay in barriers is compensated by the positive contribution in the well. Thus existence of the BHZ surface state is not possible. Similar conclusions may be reached for the hole branch, given by:

$$E_h = T + V_o = -Fk^2 + V_o \quad (6)$$

where $V_o = -M$ and $F = B + A^2/2M > 0$. Thus existence of topological surface state is incompatible with the kinetic energy definition and is not possible.